# $h_\alpha$: The Scientist as Chimpanzee or Bonobo

Loet Leydesdorff,*[a] Lutz Bornmann, [b] and Tobias Opthof[c]

**Abstract**

In a recent paper, Hirsch (2018) proposes to attribute the credit for a co-authored paper to the *α*-author—the authors with the highest *h*-index—regardless of his or her actual contribution, effectively reducing the role of the other co-authors to zero. The indicator $h_\alpha$ inherits most of the disadvantages of the *h*-index from which it is derived, but adds the normative element of reinforcing the Matthew effect in science. Using an example, we show that $h_\alpha$ can be extremely unstable. The empirical attribution of credit among co-authors is not captured by abstract models such as *h,* $\bar{h}$*,* or $h_\alpha$.

[a] *corresponding author; Amsterdam School of Communication Research (ASCoR), University of Amsterdam**,** PO Box 15793, 1001 NG Amsterdam, The Netherlands; loet@leydesdorff.net
[b] Division for Science and Innovation Studies, Administrative Headquarters of the Max Planck Society, Hofgartenstr. 8, 80539 Munich, Germany; bornmann@gv.mpg.de
[c] Experimental Cardiology Group, Heart Failure Research Center, Academic Medical Center AMC, Meibergdreef 9, 1105 AZ Amsterdam, The Netherlands; tobias.opthof@gmail.com

**Introduction**

Unlike bonobos, chimpanzees are organized in groups where the alpha male is the winner who "takes all" (de Waal, 2000 and 2006). In a recent paper (2018), the physicist Jorge E. Hirsch proposes to attribute the credit for a co-authored paper to the $α$-author, regardless of his or her actual contribution, effectively reducing the role of the other co-authors to zero. The $α$-author is defined (at p. 2) as "the co-author with the highest $h$-index." The $h$-index itself was defined by Hirsch (2005, p. 16569) as the number of papers of a scientist with at least $h$ citations. Despite its obvious shortcomings, this $h$-index has been incorporated into bibliometric databases (Web of Science, Scopus, and Google Scholar).

The $h$-index combines the number of publications and citations into a single measure that can easily be determined. Many decision makers in science prefer to have such a clear and controllable result. Furthermore, $h$-values can be attributed to all sets of publications with citations, such as departments, universities, or journals. However, the $h$-index is mathematically inconsistent (Waltman & Van Eck, 2012) and there are no convincing arguments why the numbers of publications and citations should be combined in this way; other counting rules for identifying the $h$-core among the papers are equally possible (e.g., Egghe, 2006; Ye, 2017).

In response to the critique that the $h$-index does not take the number of co-authors of a paper into account, Hirsch (2010) extended his original $h$-index with $\bar{h}$ as follows: "A scientist has index $\bar{h}$ if $\bar{h}$ of his/her papers belong to his/her $\bar{h}$-core. A paper belongs to the $\bar{h}$ core of a scientist if it has $≥\bar{h}$ citations and in addition belongs to the $\bar{h}$-core of each of the co-authors of the paper" (p.

742). The contribution to the $\bar{h}$ of an individual scientist is thus made dependent on the achievements of his/her co-authors. The newly proposed $h_\alpha$ dissolves this dependency and focuses exclusively on the seniority ("leadership") of the individual scientist. But it also generates some new problems.

The numbers of citations of each of the co-authors at the moment of publication are not retrievable in the bibliometric databases at later moments. As in the case of $\bar{h}$, Hirsch makes a pragmatic concession for the operationalization by using the number of papers in the current *h*-core of the scientist as a proxy. The $h_\alpha$ can then be obtained as follows: "One simply has to go through the list of papers in the *h*-core of a scientist and eliminate those papers for which a coauthor has higher *h*-index than the *h*-index of the author under consideration" (Hirsch, 2018, at p. 2). Division of this value of $h_\alpha$ by the *h*-value provides a ratio $r_\alpha = (h_\alpha / h)$ between zero and one which can also be expressed as a percentage.

The consequences of this operationalization are devastating for the value of $h_\alpha$. For example: let us assume that three authors (A, B, and C) share an *h*-core of 50 papers. The papers are cited from 110 times for paper #1 and one time less for each next paper, ending at 61 citations for paper #50. Therefore, both their *h*-index and their $h_\alpha$ index are 50. Additionally, each of the three authors has a single-authored paper which is cited 49 times. This is the situation at time *t*. Two months later, however, A's single-authored paper receives two new citations, bringing her *h*-index as well as her $h_\alpha$ index to 51. At that very moment, the $h_\alpha$ for authors B and C decrease from 50 to zero as a consequence of the citation of a paper which may have no relation to the collaboration among these three authors. One month later, the single-authored paper of B also receives two additional citations: the $h_α$-index of B increases from 0 to 51, but the $h_α$ of C remains zero. However, the *h*-indices of the three authors are 51, 51, and 50, respectively. The $h_α$, however, can be extremely unstable.

**Analytical and normative use and assessment**

In our opinion, indicators such as *h*, $\bar{h}$, or $h_α$ (and the many *h*-index variants proposed hitherto; see Bornmann, Mutz, Hug, & Daniel, 2011) can be evaluated (*i*) analytically and empirically as a methodology in bibliometrics and science studies, and (ii) normatively as an indicator providing management information. The *h*-index itself, for example, has virtually no analytical value, as has been shown extensively in the scientometric literature (e.g., Bornmann, 2014), but it is frequently used in research management and by policy-makers. Normatively successful indicators can function performatively in competitive environments (Dahler-Larsen, 2014). For example, indicators can be incorporated into bureaucratic processes and function then as institutional incentives (Wouters, 2014). Applicants, for example, nowadays routinely report their *h*-index.

The newly proposed indicator $h_α$ inherits most of the disadvantages of the *h*-index from which it is derived (e.g., Marchant, 2009), but adds the normative element of reinforcing the Matthew effect in science, which was defined by Merton (1968) based on the following passage from the Gospel: "For unto every one that hath shall be given, and he shall have abundance: but from him that hath not shall be taken away even that which he hath" (Matthew 25:29, King James version).

This tendency will prevail in some sciences more than others, but it can be reinforced by using the $h_\alpha$ for the attribution of credit, implying that "the winner takes all."

However, Hirsch's models do not describe the attribution of credit in empirical situations. The literature informs us that the attribution of credit differs among the disciplines (e.g., Moed, 2000; Price, 1970; Wagner, 2008). The order of authorship in the byline of the article is accordingly pluriform. In the life sciences, for example, papers are often attributed to the PhD student or postdoc as the first author and to the supervisor as the last one, while in economics the names of co-authors are commonly listed in alphabetical order. A senior with the largest *h*-value may also be involved, but not necessarily in one of these two (junior or senior) functions; perhaps, for legitimatory purposes or in relation to funding agencies. In other words, the empirical attribution of credit among co-authors is not captured by abstract models such as $\bar{h}$ or $h_\alpha$.

Evaluation using publication and citation measures should consider the field-specific environments in which the evaluated scientists operate and the objectives of the evaluation: are research groups in biomedicine being compared, or candidates for a full professorship in economics? Bornmann & Marewski (2018) introduced the term "bibliometrics-based heuristics," which emphasizes the meaning of the environment in which the evaluation takes place. One cannot make performance judgements without information about the international network of the evaluees, the quality of the journals in which their papers were published, the number of single-authored papers compared to the number of co-authored papers, the concrete topics of the scientists' research, and the most important papers in their careers.

If, for other reasons, a single number is needed that reflects both impact and output dimensions in a comparison, the number of papers which belong to the 10% most frequently cited in the corresponding fields and publication years is probably the best candidate (Leydesdorff, Bornmann, Mutz, & Opthof, 2011; Narin, 1987; Tijssen, Visser, & Van Leeuwen, 2002). An age-normalized variant of this indicator (at the individual level) can be obtained by dividing this number by the years since publishing one's first paper (Bornmann & Marx, 2014).

As against these bibliometric indicators, $h$, $\bar{h}$, and $h_α$ are mathematical constructs which are formal and thus devoid of meaning. A mathematical model of how to combine publication and citation analysis without empirical testing and theoretical backing tells us more about the imagination than about the modeled system. Scientists, for example, could be scaled on behaving like chimpanzees or bonobos, and one could design a research project testing the differences in α-behavior among the disciplines. The current proposal of $h_α$, however, claims validity across the disciplines but is both untestable and uninformed; it provides us rather with a perspective. Is this, perhaps, the perspective "which forces a man to become a physicist" (Leydesdorff & van Erkelens, 1981; Mittrof, 1974)?